\documentstyle[amssymb,preprint,aps]{revtex}

\begin{document}
\title{Contextual, Optimal and Universal Realization of\ the Quantum Cloning
Machine and of the NOT gate}
\author{Francesco De Martini, Daniele Pelliccia, and Fabio Sciarrino}
\address{Dipartimento di Fisica and \\
Istituto Nazionale per la Fisica della Materia\\
Universit\`{a} di Roma ''La Sapienza'', Roma, 00185 - Italy}
\maketitle

\begin{abstract}
A simultaneous realization of the Universal Optimal Quantum Cloning Machine
(UOQCM)\ and of the Universal-NOT gate by a quantum injected optical
parametric amplification (QIOPA), is reported. The two processes, forbidden
in their{\it \ exact} form for fundamental quantum limitations, are found 
{\it universal} and {\it optimal, }and the measured{\it \ fidelity F {\tt 
\mbox{$<$}%
} 1 }is found close to the limit values evaluated by quantum theory. This
work may enlighten the yet little explored interconnections of fundamental
axiomatic properties within the deep structure of quantum mechanics.
\end{abstract}


A most interesting aspects of quantum information consists of its insightful
enlightening the deep structure of quantum mechanics, the cornerstone of our
yet uncertain understanding of the Universe. Because of its key role in
various branches of science and of the immediate confrontation with its well
alive ''classical'' counterpart, Quantum Information (QI) has indeed become
the training field for intriguing questions, apparent contradictions and
paradoxes, ''forbidden'' processes. It is then not surprising that after a
century long quantum endeavour the inner structure of the theory has been
drawn recently to a profound investigation. For instance, several unexpected
quantum ''bounds'' were discovered mostly in quantum measurement theory. A
renowned example is the ''no-cloning'' theorem implying the impossibility of
using of the Einstein-Podolsky-Rosen (EPR) correlations for the superluminal
communication of significant messages \cite{1,2,3}. Another important,
recently discovered forbidden QI process consists of the impossibility of
realizing in a 2-dimensional Hilbert space the \ Universal NOT-gate, i.e. a
spin-flip map by which {\it any} input qubit $\left| \Psi \right\rangle $ is
transformed into the corresponding orthogonal qubit $\left| \Psi ^{\perp
}\right\rangle $ \cite{4}. In the present work these two forbidden processes
are realized for the first time ''contextually'' and ''optimally'', i.e.
with the maximum approximation allowed by QM, by two separate entangled
parts of the same apparatus: a Quantum Injected Optical Parametric Amplifier
(QIOPA) \cite{5}. In analogy with the original definition by Kochen and
Specker (KS) here ''contextuality'' implies the inter-dependence of \ local
measurements on distant systems connected by a bipartite entanglement \cite
{6,7}. An unexpected and quite intriguing aspect of this condition is the
fact that according to axiomatic quantum theory the ''impossibility'' of \
two forbidden processes lies on the violation of two {\it distinct} and
independent necessary QM requirements, namely the {\it linearity} and the 
{\it complete} {\it positivity} (CP) of any QM map \cite{6}. There is
evidence that this interesting correlation reflects a very general property
of the optimal quantum transposition map in {\it any} entangled system.

The two processes were realized simultaneously in a $2\times 2\dim $ensional
Hilbert space ${\it H}_{1}\otimes {\it H}_{2}$ of photon polarization $(%
\overrightarrow{\pi })${\bf \ }and the linearized 1-to-2 cloning process,
i.e. leading from $N=1$ input qubit to $M=2$ clones, was investigated.
Consider first the case of an input $\overrightarrow{\pi }$-encoded qubit $%
\left| \Psi \right\rangle _{in}\ $associated with a single photon with
wavelength (wl) $\lambda $, injected on the input mode $k_{1}$of the QIOPA,
the other input mode $k_{2}$ being in the vacuum state \cite{4}. As for
previous works, the photon was injected into the a nonlinear (NL) BBO ($%
\beta $-barium-borate) 1.5 mm thick crystal slab, cut for Type II phase
matching and excited by a sequence of UV\ mode-locked laser pulses having
duration $\tau \approx $140 $f\sec $ and wl. $\lambda _{p}$:\ Figure 1. The
relevant modes of the NL 3-wave interaction driven by the UV pulses
associated with mode $k_{p}$ were the two spatial modes with wave-vector
(wv) $k_{i}$, $i=1,2$, each supporting the two horizontal $(H)$ and vertical 
$(V)$ {\it linear}-$\overrightarrow{\pi }$'s of the interacting photons. The
QIOPA was $\lambda $-degenerate, i.e. the interacting photons had the same
wl's $\lambda =%
{\frac12}%
\lambda _{p}=795nm$. The NL\ crystal orientation was set as to realize the
insensitivity of the amplification quantum efficiency $(QE)\;$to any input
state $\left| \Psi \right\rangle _{in}$, i.e. the {\it universality} (U)\ of
the ''cloning machine'' and of the U-NOT gate under investigation.$\ \ $It\
is well\ known that this key property is assured by the squeezing
hamiltonian \cite{4,8}: 
\begin{equation}
\widehat{H}_{int}=i\chi \hbar \left( \widehat{a}_{\Psi }^{\dagger }\widehat{b%
}_{\Psi \perp }^{\dagger }-\widehat{a}_{\Psi \perp }^{\dagger }\widehat{b}%
_{\Psi }^{\dagger }\right) +h.c.  \label{1}
\end{equation}
The field operators sets $\left\{ \widehat{a}_{\Psi }^{\dagger },\widehat{a}%
_{\Psi }\right\} $,$\left\{ \widehat{a}_{\Psi \perp }^{\dagger },\widehat{a}%
_{\Psi \perp }\right\} $,$\left\{ \widehat{b}_{\Psi }^{\dagger },\widehat{b}%
_{\Psi }\right\} $,$\left\{ \widehat{b}_{\Psi \perp }^{\dagger },\widehat{b}%
_{\Psi \perp }\right\} $refer to two mutually orthogonal $\overrightarrow{%
\pi }$-states, $\left| \Psi \right\rangle $ and $\left| \Psi ^{\perp
}\right\rangle $, realized on the two interacting spatial modes $k_{i}$. The 
$\widehat{a}$ and $\widehat{b}\ $operators refer to modes $k_{1}$ and $k_{2}$%
, respectively \cite{4}. The $SU(2)$ invariance of $\widehat{H}_{int}$
implied by the U condition, i.e. the independence of the OPA ``gain'' $%
g\equiv \chi t$ to any unknown $\overrightarrow{\pi }${\bf - }state of the
injected qubit, $t$ being the interaction time, allows the use of the
subscripts $\Psi $ and $\Psi ^{\perp }$ in Eq.1 \cite{8}.

The QIOPA apparatus adopted in the present work was arranged in the
self-injected configuration\ shown in Figure 1. The UV pump beam,
back-reflected by a spherical mirror $M_{p}$ with 100\% reflectivity and $%
\mu -$adjustable position ${\bf Z}$, excited the NL crystal in both
directions $-k_{p}$ and $k_{p}$, i.e. correspondingly oriented towards the
right hand side and the l.h.s. of Fig.1. A Spontaneous Parametric Down
Conversion (SPDC) process excited by the $-k_{p}$ UV\ mode created {\it %
singlet-states} of photon polarization $(\overrightarrow{\pi })$. The photon
of each SPDC pair emitted over $-k_{1}$ was back-reflected by a spherical
mirror $M$ into the NL crystal and provided the $N=1$ {\it quantum injection}
into the OPA excited by the UV beam associated with the back-reflected mode $%
k_{p}$. Because of the low pump intensity, the probability of the unwanted $%
N=2$ injection has been evaluated to be $10^{-2}$ smaller than the one for $%
N=1$. The twin SPDC\ photon emitted over mode $-k_{2}$ , selected by the
devices (Wave-Plate + Polarizing Beam Splitter: $WP_{T}\ $+ $PBS_{T}$) and
detected by $D_{T}$, provided the ''trigger'' of the overall conditional
experiment. Because of the EPR non-locality of the emitted singlet, the $%
\overrightarrow{\pi }$-selection made on $-k_{2}$ implied deterministically
the selection of the input state $\left| \Psi \right\rangle _{in}$on the
injection mode $k_{1}$. By adopting $\lambda /2$ or $\lambda /4$ wave-plates
(WP) with different orientations of the optical axis, the following $\left|
\Psi \right\rangle _{in}$states were injected:$\;\left| H\right\rangle $, $%
2^{-1/2}(\left| H\right\rangle +\left| V\right\rangle )$, and $%
2^{-1/2}(\left| H\right\rangle +i\left| V\right\rangle )$. The three fixed
quartz plates $(Q)$ inserted on the modes $k_{1}$, $k_{2}$ and $-k_{2}$
provided the compensation for the unwanted walk-off effects due to the
birefringence of the NL crystal. An additional walk-off compensation into
the BBO\ crystal was provided by the $\lambda /4$ WP exchanging on mode $%
-k_{1}\ $the $\left| H\right\rangle $ and $\left| V\right\rangle \ 
\overrightarrow{\pi }-$ components of the injected photon. All adopted
photodetectors $(D)$ were equal SPCM-AQR14 Si-avalanche single photon units
with $QE^{\prime }s\cong 0.55$. One interference filter with bandwidth $%
\Delta \lambda =6nm$ was placed in front of each $D$.

Let us consider the injected photon in the mode $k_{1}$ to have any
polarization $\overrightarrow{\pi }{\bf =}\Psi $, corresponding to the
unknown input qubit $\left| \Psi \right\rangle _{in}$. We express this $%
\overrightarrow{\pi }$-state as $\widehat{a}_{\Psi }^{\dagger }\left|
0,0\right\rangle _{k_{1}}=\left| 1,0\right\rangle _{k_{1}}$ where $\left|
m,n\right\rangle _{k_{1}}$ represents a product state with $m$ photons of
the mode $k_{1}$ having the polarization $\Psi $, and $n$ photons having the
polarization $\Psi ^{\perp }$. Assume the input mode $k_{2}$ to be in the 
{\it vacuum state}. The initial $\overrightarrow{\pi }$-state of modes $k_{i}
$ reads $\left| \Psi \right\rangle _{in}=\left| 1,0\right\rangle
_{k_{1}}\otimes \left| 0,0\right\rangle _{k_{2}}$ and evolves according the
unitary operator $\widehat{{\bf U}}\equiv \exp \left( -i\widehat{H}%
_{int}t\right) $: 
\begin{equation}
\widehat{{\bf U}}\left| \Psi \right\rangle _{in}\simeq \left|
1,0\right\rangle _{k_{1}}\otimes \left| 0,0\right\rangle _{k_{2}}+g\left( 
\sqrt{2}\left| 2,0\right\rangle _{k_{1}}\otimes \left| 0,1\right\rangle
_{k_{2}}-\left| 1,1\right\rangle _{k_{1}}\otimes \left| 1,0\right\rangle
_{k_{2}}\right) 
\end{equation}
This represent the 1st-order approximation for the${\Bbb \ }$pure output
state vector $\left| \Psi \right\rangle _{out}\ $for $t>0$ \cite{4}. Be $%
\varrho \equiv $ $(\left| \Psi \right\rangle \left\langle \Psi \right|
)_{out}$ the corresponding density operator. It has been shown that the
amplified term $\varpropto g$ in $\left| \Psi \right\rangle _{out}=\widehat{%
{\bf U}}\ \left| \Psi \right\rangle _{in}$ is equal to the output state $%
\left| \Psi _{N}\right\rangle _{M}$ of a general Universal Optimal Quantum
Cloning Machine (UOQCM),$\ $where $N$ and $M$ $>N\ $are respectively the
number of input qubits to be cloned and the number of clones \cite{8,9}. The
above linearization procedure, i.e. the restriction to the simplest case $N=1
$, $M=2$, is justified here by the small experimental value of the {\it gain}%
: $g\approx 0.1$. The zero order term in Eq.2 corresponds to absence of NL
interaction while the second term describes the 1st-order QIOPA\
amplification process. In this context, the state $\left| 2,0\right\rangle
_{k_{1}}$ expressing two photons of the mode $k_{1}$ in the $\overrightarrow{%
\pi }$-state $\Psi $ corresponds to the state $\left| \Psi \Psi
\right\rangle $ expressed by the general theory and implies the $M=2$
cloning of the input $N=1$ qubit \cite{1,9,10}.\ Contextually with the
realization of cloning on mode $k_{1}$,\ the vector $\left| 0,1\right\rangle
_{k_{2}}$ in Eq.2 expresses the single photon state on mode $k_{2}$ with
polarization $\Psi ^{\perp }$, i.e. the {\it flipped} version of the input
qubit. Then the QIOPA acts simultaneously on the output mode $k_{1}$ as an
approximate UOQCM and on the output mode $k_{2}\ $as an approximate
Universal NOT-gate, i.e. two processes which are forbidden in their {\it %
exact} form, as said \cite{4}.

To see that the stimulated emission is indeed responsible for creation of
the flipped qubit, let us compare the state Eq.2 with the output of the OPA
when the {\it vacuum} {\it state} is injected into the NL crystal on {\it %
both} input modes $k_{i}$. In this case is $\left| \Psi \right\rangle
_{in}^{0}=\left| 0,0\right\rangle _{k_{1}}\otimes \left| 0,0\right\rangle
_{k_{2}}$ and we obtain to the 1st order of approximation: $\widehat{{\bf U}}%
\left| \Psi \right\rangle _{in}^{0}\simeq \left| 0,0\right\rangle
_{k_{1}}\otimes \left| 0,0\right\rangle _{k_{2}}+g\left( \left|
1,0\right\rangle _{k_{1}}\otimes \left| 0,1\right\rangle _{k_{2}}-\left|
0,1\right\rangle _{k_{1}}\otimes \left| 1,0\right\rangle _{k_{2}}\right) $.
We see that the flipped qubit expressed by the state $\left|
0,1\right\rangle _{k_{2}}$ in the r.h. sides of \ the last equation and of \
Eq. 2 appears with different amplitudes corresponding to the ratio of
probabilities to be equal to $R^{\ast }=2:1$. The quantity $R^{\ast }$\ may
be referred to as ''signal-to-noise'' ratio: $S/N$.\ \ Note also in these
equations that, by calling by $R$ \ the ratio of the probabilities of
detecting $2$ and $1$ photons on mode $k_{1}$, we obtain: $R=R^{\ast }$. In
other words, and highly remarkably the {\it same value} of $S/N$ affects
both cloning and U-NOT processes realized contextually on the two different
output modes: $k_{1}$and $k_{2}$. The ratios $R$ \ and $R^{\ast }$ are
indeed the quantities measured in the present experiments for both UOQCM and
U NOT-gate processes, respectively. These ratios are used to determine the
corresponding values of the {\it fidelity,} defined as follows \cite{4,9,10}%
: {\it Cloning} {\it fidelity}$:$ $F=Tr(\rho _{1}\widehat{n}_{1\pi
})/Tr(\rho _{1}\widehat{n}_{1})=(2R+1)/(2R+2)$. U-NOT {\it fidelity}$:${\it %
\ }$F^{\ast }=Tr(\rho _{2}\widehat{n}_{2\bot })/Tr(\rho _{2}\widehat{n}%
_{2})=R^{\ast }/(R^{\ast }+1)$. There:$\widehat{\ n}_{i}=\widehat{n}_{i\Psi
}+\widehat{n}_{i\bot }$, being $\widehat{n}_{1\Psi }\equiv \hat{a}_{\Psi
}^{\dagger }\hat{a}_{\Psi }$, $\widehat{n}_{1\bot }\equiv \hat{a}_{\Psi \bot
}^{\dagger }\hat{a}_{\Psi \bot }$,$\widehat{n}_{2\Psi }\equiv \widehat{b}%
_{\Psi }^{\dagger }\widehat{b}_{\Psi }$, $\widehat{n}_{2\bot }\equiv 
\widehat{b}_{\pi \bot }^{\dagger }\widehat{b}_{\pi \bot }$ and the reduced
density operators of the output state on the modes $k_{i}$ are: $\varrho
_{1}=Tr_{2}\rho $ and $\varrho _{2}=Tr_{1}\rho $. The values $F=5/6$ and $%
F^{\ast }$ $=2/3$ are the ''optimal'' values corresponding to the limit
values $R=R^{\ast }=2$ allowed by QM, respectively for cloning $N=1$ into $%
M=2$ qubits and for realizing a U-NOT gate by single qubit flipping.

The goal of the cloning experiment was to measure, under injection of the
state $\left| \Psi \right\rangle _{in}$, the $S/N$ related to the OPA\
amplification leading to the state $\left| 2,0\right\rangle _{k_{1}}$on the
output mode $k_{1}$, here referred to as the ''cloning mode'' (C). At the
same time on the C mode no amplification should affect the output state $%
\left| 1,1\right\rangle _{k_{1}}$ corresponding to $\overrightarrow{\pi }%
^{\perp }$ orthogonal\ to $\overrightarrow{\pi }$. In order to perform this
task, the $PBS_{2}$ was removed on the mode $k_{2}$ and the photons on the
same mode detected by a single detector: $D_{2}$. The output $M=2$ photons
associated with\ the C mode were separated by means of a $50:50$ {\it %
conventional }Beam Splitter $(BS)$\ and their $\overrightarrow{\pi }$-states
analyzed by the combinations of $WP_{1}$ and of $PBS_{1a}$ and $PBS_{1b}$.
For each injected $\overrightarrow{\pi }$-state, $\left| \Psi \right\rangle
_{in}$ $WP_{1}$ was set in order to detect $\left| \Psi \right\rangle $ by $%
D_{a}$ and $D_{b}$ and to detect $\left| \Psi ^{\perp }\right\rangle $,
orthogonal to $\left| \Psi \right\rangle $ by $D_{b}^{\ast }$. Hence any
coincidence event detected by $D_{a}$ and $D_{b}$ corresponded to the
realization of the state $\left| \Psi \Psi \right\rangle _{1}$ over the C
mode, while a coincidence detected by $D_{a}$ and $D_{b}^{\ast }$
corresponded to the state $\left| \Psi \Psi ^{\perp }\right\rangle _{1}$.

The measurement of $R$ could be carried out by 4-coincidence measurements
involving simultaneously the detector sets: $[D_{2},D_{T},D_{a},D_{b}]$ and $%
[D_{2},D_{T},D_{a},D_{b}^{\ast }]$:\ $R=%
{\frac12}%
C_{1}/C_{2}$ being $C_{1}$ and $C_{2}$ the results obtained by the sets,
respectively. A better alternative method, actually adopted in the
experiment is described in: \cite{11}.

The experimental data reported\ on the l.h.s. column of Figure 2 correspond
to the 4-coincidence measurement by $[D_{2},D_{T},D_{a},D_{b}]$ that is, to
the emission over the C-mode of the ''cloned'' state $\left| \Psi \Psi
\right\rangle _{1}$ under injection of the state $\left| \Psi \right\rangle
_{in}$. The resonance peaks found by this measurement identified the
position ${\bf Z}$ of the UV mirror $M_{p}$ corresponding to the maximum
overlapping of the pump and of the injected pulses, i.e. to the actual
realization of the QIOPA operation. According with the analysis above the
''noise'' plots, implying the realization of the state $\left| \Psi \Psi
^{\perp }\right\rangle _{1}$measured by the 4-coincidence set $%
[D_{2},D_{T},D_{a},D_{b}^{\ast }]$, do no show any OPA\ amplification
effect. Each pair of adjacent plots belonging to the two columns in Fig.2
corresponded to the same injected $\overrightarrow{\pi }$-state $\left| \Psi
\right\rangle _{in}$. \ Precisely, starting from the upper pair down, {\it %
linear-}$\overrightarrow{\pi }$ {\it horizontal}, {\it linear-}$%
\overrightarrow{\pi }$ at $45%
{{}^\circ}%
$, {\it left} {\it circular}-$\overrightarrow{\pi }$. The corresponding
experimental values of the cloning fidelity were found: $F_{H}=0.812\pm
0.007 $;$\ F_{H+V}=0.812\pm 0.006$; $F_{left}=0.800\pm 0.007$, to be
compared with the {\it optimal} values$\ F_{th}=5/6\approx 0.833$
corresponding to the limit value: $R=2$. This result is consistent with the
one obtained by A. Lamas-Linares {\it et al.} by adoption of \ a {\it %
semi-classical, coherent} injected field \cite{10}.

The U-NOT\ gate operation has been demonstrated by restoring the $PBS_{2}$
on the mode $k_{2}$, the ''anti-cloned'' mode (AC), coupled to the detectors 
$D_{2}$ and $D_{2^{\ast }}$, via the $WP_{2}$, as shown in Fig1. The $%
\overrightarrow{\pi }\;$-analyzer consisting of $(PBS_{2}+WP_{2})$ was set
as to reproduce the {\it same} filtering action of the analyzer $%
(PBS_{T}+WP_{T})$ for the ''trigger'' signal. The devices $PBS_{1a}$ and $%
PBS_{1b}$ were removed on the C channel and the field was simply coupled by $%
BS_{1}\;$to the detectors $D_{a}$ and $D_{b}$. A coincidence event involving
these ones was the {\it signature} for a {\it cloning} event. The values of
the $S/N$ ratio $R^{\ast }$ measured by 4-coincidence experiments involving
the sets $[D_{2},D_{T},D_{a},D_{b}]$ and $[D_{2^{\ast }},D_{T},D_{a},D_{b}]$
and reported in the r.h.s column in Fig 2, were adopted to determine the
values of the U-NOT {\it fidelity} $F^{\ast }$. The results are: $%
F_{H}^{\ast }=0.630\pm 0.008$;$\ F_{H+V}^{\ast }=0.625\pm 0.016$; $%
F_{left}^{\ast }=0.618\pm 0.013$ to be compared with the {\it optimal}
value: $F^{\ast }=2/3\approx 0.666\ $\cite{11}. Note that all results
reported in Fig. 2 show an amplification efficiency which is almost
identical for the different input qubits:\ $\left| \Psi \right\rangle _{in}$%
. This significant result represents the first demonstration of the {\it %
universality} of the QIOPA\ system carried out by {\it quantum injection} of
a {\it single} photon qubit \cite{4,10}.

As said, the intriguing result of the present work is that both Quantum
Cloning and U-NOT\ processes are realized {\it optimally} and {\it %
contextually} by the {\it same} physical apparatus and by the {\it same}
unitary transformation over the two entangled components of a bipartite spin-%
$\frac12$%
space $H_{1}\otimes H_{2}$. To the best of our knowledge it is not well
understood yet why these {\it forbidden }processes can be so closely
related.\thinspace We may try to enlighten here at least one formal aspect
of this correlation.

Note first that the overall output vector state $\left| \Psi \right\rangle
_{out}$ expressed by Eq.\ 2 is a {\it pure state} since the unitary $%
\widehat{U}\ $acts on a pure input state. Consequently the reduced density
matrices $\rho _{1}\ $and $\rho _{2}$ have the same eigenvalues and the
entanglement of the {\it bi-partite }state $\left| \Psi \right\rangle
_{out}\ $is conveniently measured by the {\it \ entanglement entropy: }${\it %
E}(\Psi )=S(\rho _{1})=S(\rho _{2})$ being $S(\rho _{j})\equiv -Tr\rho
_{j}\log _{2}\rho _{j}$ the Von Neumann entropy of the either (C) or (AC)
subsystem, $i=1,2\ $\cite{12}. We may comment on this result by considering
first the approximate {\it cloning} process performed by the UOQCM\ acting
on the C channel, $k_{1}$. What has been actually realized in the experiment
was a procedure of \ {\it linearization} of the cloning map which is {\it %
nonlinear }and, as such{\it \ }cannot be{\it \ }realized exactly by Nature 
\cite{1,6}. By this procedure a {\it mixed-state} condition of the output
state$\ \rho _{1}$ was achieved corresponding to the {\it optimal} limit
value of the entropy $S(\rho _{1})>0$. Owing to the above expression of $%
{\it E}(\Psi )$, the {\it same} degree of {\it mixedeness }also affects{\it %
\ }the{\it \ }output state realized on the AC channel, thus verifying the
equation $R=R^{\ast }$ affecting the results shown in Fig 2. Since on the AC
channel an approximate CP map is realized which is generally distinct from
any process realized on the C channel, e.g. here the {\it cloning} process,
the above entropy equation appears to establish a significant {\it symmetry
condition} in the context of axiomatic quantum theory \cite{1}. Remarkably
enough, it has been noted recently that the transformation connecting the
cloning and U-NOT processes also realizes contextually the{\it \ optimal
entangling} process and the {\it universal probabilistic quantum processor } 
\cite{13}. Work supported by the FET European Network on Quantum Information
and Communication (Contract IST-2000-29681: ATESIT), by PRA-INFM\ ''CLON''\
and by PAIS-INFM\ 2002 (QEUPCO).

FIGURE 1.\ Schematic diagram of the {\it universal optimal cloning machine}
(UOQCM)\ realized on the cloning (C)\ channel (mode $k_{1}$) of a {\it %
self-injected} OPA and of the Universal NOT (U-NOT) gate realized on the
anticloning (AC)\ channel, $k_{2}$.

FIGURE\ 2.\ Demonstration of the UOQCM (l.h.s.) and the U-NOT gate (r.h.s.)
by 4-coincidence measurements. The values of the ''{\it fidelities}'' of the
processes $F$ and $F^{\ast }$ evaluated by each test are expressed in the
upper side of each plot. Filled squares: plots corresponding to ''correct''
polarization; Open triangles: plots corresponding to the ''noise''. The
solid lines represent the best gaussian fit to the results expressing the
''correct'' polarization.

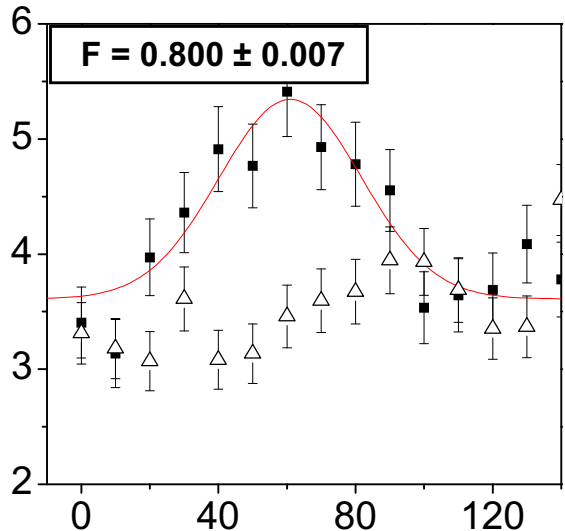
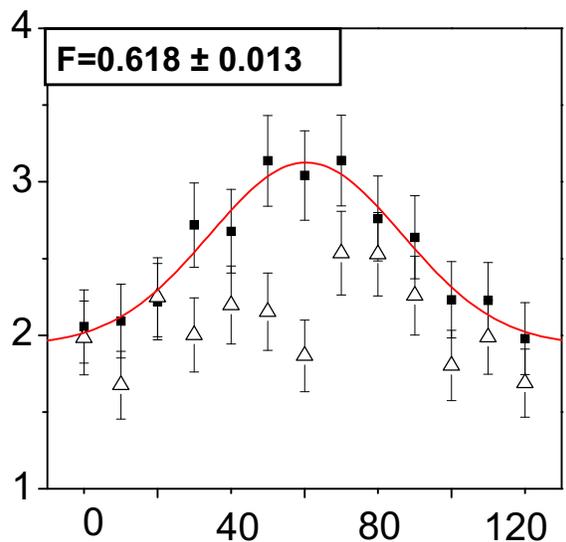
\end{document}